# Intermediate Mass Fragment Production In Symmetric Collisions

Anupriya Jain, Bahadur Singh, Suneel Kumar[#]

*School of Physics and Material Science, Thapar University, Patiala 147004, Punjab, India*
*Address*
[#]suneel.kumar@thapar.edu

*Abstract*— We present a complete systematic theoretical study of multifragmentation and its associated phenomena in heavy ion collisions. This study is performed within an Isospin dependent Quantum Molecular Dynamical Model (IQMD) and using Minimum Spanning Tree (MST) algorithm. Simulations are carried out to study the different parameters like time evolution of multiplicity, mass distribution, impact parameter dependence and IMF's production dependence of projectile & target mass. The rise and fall in multiplicity of IMF's is observed. Results are compared with experimental data of ALADIN and are found to be in close agreement.

## I. INTRODUCTION

The study of heavy-ion collisions at intermediate energies ($50 \leq A \leq 1000$) MeV/nucleon provides a rich source of information for many rare phenomena such as multifragmentation, collective flow, particle production etc. [1]. One can also shed light on the mechanism behind the fragmentation in highly excited nuclear systems. In this energy region, multifragmentation appears to be a dominant de-excitation channel apart from the other less populated channels of manifestation of liquid gas phase transition is considered as a gateway to nuclear equation of state. Numerous investigations are cited in the literature which handles the de-excitation of nuclear system in multifragmentation. The experimental analysis of the emission of intermediate mass fragments (IMFs), has yielded several interesting observations: de Souza et al. [2] showed that as the beam energy increased from 35 to 110MeV/nucleon for $^{36}$Ar+$^{197}$Au collisions, the IMF multiplicity for central collisions showed a steady increase with incident energy. Also the IMF multiplicity decreased as the collisions moved from central to peripheral. On the other hand, Tsang et al. [3], in their investigation of $^{197}$Au+$^{197}$Au from 100 to 400 MeV/nucleon, found that the IMF production peak shifted from near central towards peripheral as the beam energy was increased. For central collisions, where the excitation energy is best defined, they found a rapid decrease of the IMF multiplicity with increase in energy. A more comprehensive study was carried out by Peaslee et al. [4] in their studies of the $^{84}$Kr+$^{197}$Au from 35 to 400 MeV/nucleon, where they found that the IMF multiplicity increased with increasing energy to a maximum around 100MeV/nucleon and then decreased slowly. Stone et al. [5] used a more symmetric system of $^{86}$Kr+$^{93}$Nb from 35 to 95MeV/nucleon to obtain IMF multiplicity distribution as a function of beam energy by selecting central events. It is clear from the previous studies that for a particular system the IMF multiplicity should increase with beam energy at low energies. Competing with this trend would be the depletion of IMFs as a result of excess energy causing the IMFs to break up into smaller fragments. As the energy increases the latter phenomenon should become more dominant and the production of IMFs should decrease due to the transition into the gas phase of nuclear matter as observed by Tsang et al. Comparison of the different studies also shows that the IMF multiplicity increases with the system mass for measurements at the same energy. Our present study will shed light on multiplicity, mass distribution, impact parameter dependence and IMF's production on projectile and target mass. The present analysis is carried out within the framework of the isospin-dependent quantum- molecular dynamics (IQMD) model. Our paper is organized as follows: We briefly discuss the model in Sec. II. Our results are given in Sec. III, and we summarize the results in Sec. IV.

## II. THE MODEL

The Isospin-dependent Quantum Molecular Dynamic model (IQMD) [6] is the refinement of QMD model based on event by event method. The reaction dynamics are governed by mean field, two-body collision and Pauli blocking.
The baryons are represented by Gaussian-shaped density distributions

$$f_i(\vec{r},\vec{p},t) = \frac{1}{\pi^2\hbar^2} e^{-[\vec{r}-\vec{r}_i(t)]^2 \frac{1}{2L}} e^{-[\vec{p}-\vec{p}_i(t)]^2 \frac{2L}{\hbar^2}}$$

The successfully initialized nuclei are then boosted towards each other using Hamilton equations of motion

$$\frac{dr_i}{dt} = \frac{d\langle H\rangle}{dp_i} ; \frac{dp_i}{dt} = -\frac{d\langle H\rangle}{dr_i}$$

With $\langle H \rangle = \langle T \rangle + \langle V \rangle$ is the total Hamiltonian.

$$\langle H \rangle = \sum_i \frac{p_i^2}{2m_i} + \sum_i \sum_{j>i} \int f_i(\vec{r}, \vec{p}, t) V^{ij}(\vec{r'}, \vec{r}) \times f_j(\vec{r'}, \vec{p'}, t) d\vec{r} d\vec{r'} d\vec{p} d\vec{p'}$$

The total potential is the sum of the following specific elementary potentials [7].

$$V = V_{Sky} + V_{Yuk} + V_{Coul} + V_{mdi} + V_{loc}$$

During the propagation, two nucleons are supposed to suffer a binary collision if the distance between their centroid is

$$|r_i - r_j| \leq \sqrt{\frac{\sigma_{tot}}{\pi}}$$

Where $\sigma_{tot} = \sigma(\sqrt{s}, type)$

The collision is blocked with a possibility

$$P_{block} = 1 - (1 - P_i)(1 - P_j)$$

Fig. 1 shows change in multiplicities of IMF's for the range of impact parameters. At central collisions (b= 0 – 3 fm) the overlapping of participation zone is large and more number of IMF's are observed. As we go from central to semi central (b= 0 – 6 fm) and peripheral (b= 0 - 9 fm) the overlapping of participation zone decrease which leads to the lower number of IMF's. As clear from the fig. 1 that with increasing impact parameter the lines becomes lower and lower.

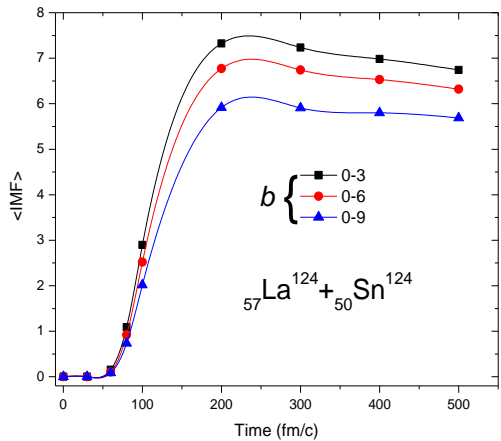

Fig.1 Time evolution of IMF's of $_{57}La^{124} + _{50}Sn^{124}$ by using range of impact parameters.

Fig.2 shows the mass distribution at three impact parameters ($\hat{b}$= 0.0, 0.4, 0.9) at 200 fm/c. As clear from the figure there is direct dependence of mass distribution on impact parameter. At central collision no heavy fragment observed because of the violent nature of central collision. In central collision no fragment observed of mass (A >20-25). In semi peripheral collisions the participant zone decreases approximately half the value so there is no complete destruction of target and projectile and fragments of A ≈ 55 observed. At peripheral collision the participant region further decrease and heavy fragments A ≈ 110 observed. The conclusion is that by increasing impact parameter the overlapping of target and projectile decreases by virtue of which heavy fragments observed. The output of heavier cluster provides a tool to determine the impact parameter.

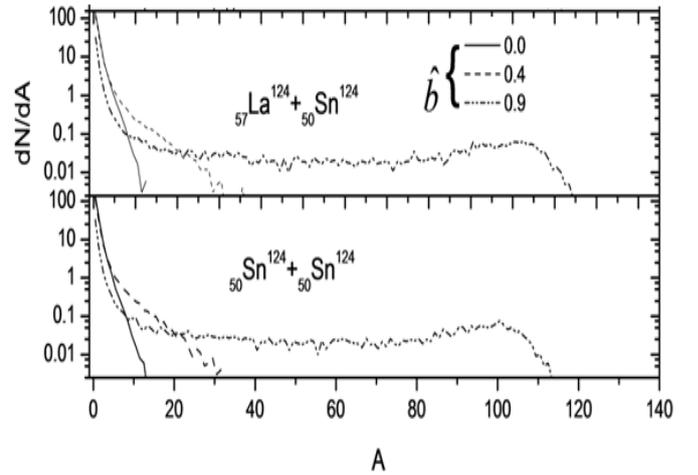

Fig.2 Mass distribution of $_{57}La^{124} + _{50}Sn^{124}$ and $_{50}Sn^{124} + _{50}Sn^{124}$

The ALADIN results are the most complete piece of data available for multifragmentation. The most prominent feature of the multifragment decay is the universality of the fragment and fragment charge correlation. The loss of memory of the entrance channel is an indication that the equilibrium is attained prerior to the fragmentation stage of the reaction. Here we are comparing our results with experimental data of reactions $_{57}La^{124} + _{50}Sn^{124}$ and $_{50}Sn^{124} + _{50}Sn^{124}$ at energy 600 MeV/nucleon [8]. Fig. 2 shows IMF's as a function of $Z_{bound}$. The quantity $Z_{bound}$ is defined as sum of all atomic charges $Z_i$ of all fragments with $Z_i > 2$. Here we observe that at semi peripheral collisions multiplicity <IMF> shows a peak because most of the spectator source does not take part in collision and large number of IMF's are observed. In case of central collision the collisions are so violent that few number of IMF's observed and for peripheral collisions very small portion of target and projectile overlap so again few number of IMF's observed most of the fragments goes out in heavy mass fragments (HMF's). In this way we get a clear 'rise and fall' in multifragmentation emission. It is observed that IMF's shows the agreement with data at low impact parameters but fails at intermediate impact parameters.

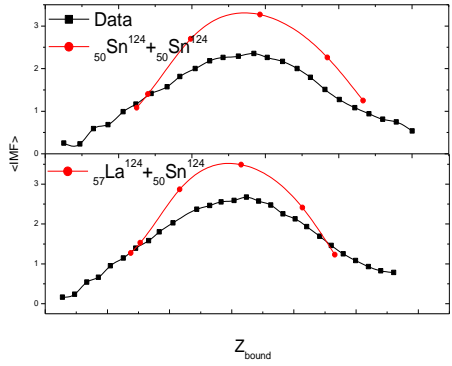

Fig. 3 Multiplicity of IMF as a function of $Z_{bound}$

## III. CONCLUSIONS

We have studied the detailed analysis of multifragmentation. Different parameters like time evolution, impact parameter dependence is studied for IMF's. It is concluded that free & LMF's have a different trend as compare to IMF's. This is due to different origin of fragments. The rise and fall in the multiplicity of IMF's is observed. The rise & fall is further compared with the experimental data of ALADIN and are found to be in close agreement. The better agreement can be obtained by taking into account different clusterization methods.

ACKNOWLEDGMENT

I owe my deepest gratitude to Dr. Rajeev K. Puri, who has been an inspiration during this research work.